\def\cS{{\cal S}}
\def\vf{\varphi}
\def\D{{\cal D}}
\def\P{{\cal P}}
\def\M{{\cal M}}
\def\l{\lambda}
\def\vf{\varphi}
\def\bm#1{\mbox{\boldmath{$#1$}}}
\def\IR{\relax{\rm I\kern-.18em R}}
\font\cmss=cmss10 \font\cmsss=cmss10 at 7pt
\def\IZ{\relax\ifmmode\mathchoice
{\hbox{\cmss Z\kern-.4em Z}}{\hbox{\cmss Z\kern-.4em Z}}
{\lower.9pt\hbox{\cmsss Z\kern-.4em Z}}
{\lower1.2pt\hbox{\cmsss Z\kern-.4em Z}}\else{\cmss Z\kern-.4em Z}\fi}

 \documentstyle[sprocl,epsf]{article}
\begin{document}
 \title{RELATIVE ENTROPY IN FIELD THEORY, THE $H$ THEOREM AND THE
RENORMALIZATION GROUP}
 \author{ J. GAITE }
 \address{Instituto de Matem{\'a}ticas y F{\'\i}sica Fundamental, 
             C.S.I.C., Serrano 123,\\ 28006 Madrid, Spain}
 %%%%%%%%%%%%%%%%%%%%%%%%%%%%%%%%%%%%%%%%%%%%%%%%%%%%%%%%%%%%%
 %You may repeat \author \address as often as necessary
 %%%%%%%%%%%%%%%%%%%%%%%%%%%%%%%%%%%%%%%%%%%%%%%%%%%%%%%%%%%%%
 \maketitle\abstracts{We consider relative entropy in Field Theory as a
well defined (non-divergent) quantity of interest. We establish 
a monotonicity property with respect to the couplings in the
theory. As a consequence, 
the relative entropy in a field theory with a
hierarchy of renormalization group fixed points ranks the fixed points
in decreasing order of criticality.
We argue from a generalized $H$ theorem that
Wilsonian RG flows induce an increase in entropy and propose
the relative entropy as the natural quantity which increases
from one fixed point to another in more than two dimensions.
  }

\section{Introduction}

This work is motivated by the old idea of irreversibility in 
the renormalization group (RG). Specifically, the existence of a
function monotonic with the RG (a Lyapunov function for it) was proposed to
ensure the existence of regular fixed points as opposed to other
possible pathological behavior, such as limit cycles.\cite{WZ} As a
sufficient condition for the existence of that function, one might have
that the RG is actually a gradient dynamical system, that is, it derives
from a potential, which is then the monotonic function. This type of RG
flow would have further convenient properties. For example, it
implies that the matrix that represents the linear RG near a fixed point
(FP) is symmetric and hence it has real eigenvalues, to be identified
with anomalous dimensions. Despite evidence---only
perturbative and at low order---for the gradient flow,\cite{WZ} its
existence remains inconclusive. 

However, in two dimensions a monotonic function has been found with
the methods of conformal field theory, Zamolodchikov's $C$
function. There have been many attempts to generalize the $C$ theorem to
higher dimensions but they cannot be considered 
definitive.\cite{Cardy,Osborn,CFL} 
Moreover, these authors have ignored Zamolodchikov's
initial motivation, the adaptation of the renowned Boltzmann's $H$
theorem to the RG setting. In a synthetic formulation we may quote this theorem
as the statement that non-equilibrium (coarse-grained) entropy
increases with time in the approach to equilibrium. Kadanoff block
transformations as well as other coarse-graining formulations of the RG
discard short-range information on a physical system and hence seem
irreversible in a similar sense to the situations described by the $H$
theorem, where the physical time is replaced by the renormalization
group parameter.

In this spirit, we define a field theoretic entropy and study its
properties, in particular, its possible divergences and its monotinicity. 
In a phase diagram with a multicritical point 
the entropy relative to it is finite and a monotonic 
function of the coupling constants spanning that diagram measured from 
that point. Therefore, this entropy is monotonic in the crossover from 
this point to another more stable multicritical point. It 
can be realized by a RG flow in which one or several relevant 
coupling constants 
become irrelevant. This is the typical situation that arises in field 
theoretic formulations of the RG. However, in coarse-grained or Wilson 
formulations the number of coupling constants is very large, as
a consequence of the action of the RG itself, and they are classified
as relevant or irrelevant only near the end fixed point. In analogy with
the statistical mechanics of irreversible processes, we can then invoke
a very general formulation of the $H$ theorem due to E.T.\ Jaynes to
show that the entropy increases in this more general context as well.  

This account is based on previous work on this subject.\cite{DI}

\section{Field theory entropy. Definition and properties}

Let us begin by recalling some concepts of probability theory. Given a
probability distribution $\{p_m\}$ one can define an entropy
\begin{equation}
S(\{p_m\}) = -\sum_{m=1}^M p_m \log p_m.
\label{entropy}
\end{equation}
For a continous distribution
\begin{equation}
S[p(x)] = -\int dx\, p(x) \log p(x).
\end{equation}
In this case is more a useful to define relative entropy with
reference to a given distribution $q(x)$
\begin{equation}
S[p(x)] = \int dx\, p(x) \log {p(x)\over q(x)}.
\end{equation}
If $q(x)$ is the uniform distribution this definition coincides up to a
conventional sign and a constant with the
previous one, now called the absolute entropy. However, it is more
correct to take as reference a normalizable distribution, for example,
the Gaussian distribution. 

Now we are ready to define an entropy in field theory. We begin with the
usual definition of the generating functional or partition function
\begin{equation}
Z[\{\l\}]= e^{-W[\{\l\}]} =
\int\D[\phi]\,{\rm e}^{-I[\phi,\{\l\}]}
\end{equation}
with $W[\{\l\}]$ the generator of connected amplitudes. 
If we write
\begin{equation}
\int\D[\phi]\,{\rm e}^{-I[\phi,\{\l\}]+W[\{\l\}]} = 1,
\end{equation}
we can consider $$\P[\phi,\{\l\}] = {\rm
e}^{-I[\phi,\{\l\}]+W[\{\l\}]}$$ as a (family of) probability 
distribution of the stochastic variables $\phi$. Therefore, the relative
entropy is 
\begin{eqnarray}
\cS[\P,\P_0]={\rm Tr}[\P\ln(\P/\P_0)] \nonumber\\
=\int\D[\phi]\,(-I+I_0+W-W_0)\,e^{-I+W} = W-W_0 - \langle I-I_0\rangle
\label{relS}
\end{eqnarray}
with $I_0$ a reference action of the same family, that is, with given
values of the coupling constants. 

The typical form of the action is a linear combination of composite
fields with their respective coupling constants,
\begin{equation}
I[\phi,\{\l\}]=\l^af_a[\phi].
\end{equation}
We assume that only a part of the action $I_r$ is of interest to us.
(We shall call it later relevant o crossover action for physical
reasons.) We
introduce a global coupling constant $z$ for it or, in other words, we
extract a common factor from the coupling constants in it,
\begin{equation}
z\,I_r = I - I_0.
\end{equation}
Then we can express the relative entropy as the Legendre transform of
$W-W_0$ with respect to $z$:
\begin{equation}
\cS_{\rm rel} = W(z) - W(0) - z{dW\over dz}. \label{Legtrans}
\end{equation}
Alternatively, it is the Legendre transform with respect to the
couplings in $I_r$, the relevant couplings $l_a$,
\begin{equation}
\cS_{\rm rel} = W - W_0 - l^a\partial_a W. \label{Legtrans2}
\end{equation}
In turn the absolute entropy is the Legendre transform with respect to
all the couplings. Both entropies have a simple description in
perturbation theory: They are sums of n particle irreducible (nPI) 
diagrams with external composite fields at zero momentum. 

We have already remarked that for continuous probability distributions
the relative entropy is well defined whereas the absolute entropy is
not. In Field Theory, however, the stochastic variable is a field or, in
other words, there is an infinity of stochastic variables that causes
that the previous quantities are not necessarily well defined. This is
the well known problem of divergences, either ultraviolet (UV) or
infrared (IR). If we assume that the field theory under study is
regularized by defining it on a finite lattice, we return to the case
of a finite number of (continuous) stochastic variables and $W$ as well
as the relative entropy are well defined. Then the first limit we have
to consider is the infinite volume or thermodynamic limit, as is usual
in Statistical Mechanics. To perform the limit one has to define
specific quantities---per lattice site or unit volume---that remain
finite. Nevertheless, at the critical point some of them may diverge, 
as occurs to the specific heat. We speak of an IR divergence. 
In (\ref{relS}) or (\ref{Legtrans2}) the possible IR divergences of 
the relative entropy are in the sum of $\langle f_a\rangle$ and hence it 
is IR finite in $d>2$.

The second and more important limit is the continuum limit, which allows
to define a field theory as the number of lattice points per unit
volume goes to infinity. Since $W$ per lattice site is finite, the
relevant quantity now, $W$ per unit volume, diverges as $a^{-d}$, with
$a$ the lattice spacing and $d$ the dimension. This is a UV divergence, 
closely related with 
the divergent ground state energy in Quantum Field Theory, which is 
substracted by taking the composite operators normal ordered. Fortunately,
the $a^{-d}$ divergence of $W$ is cancelled in the Legendre transform 
(\ref{Legtrans}) that gives the relative entropy by substracting $W(0)$. 
Furthermore, the next-to-leading divergence $a^{-d+2}$ cancels in 
the substraction of $z{dW\over dz}$. Therefore, in dimension $d<4$ 
the relative entropy is UV finite. Notwithstanding this, we know that 
if we calculate nPI diagrams in perturbation theory we shall find
UV divergences. These divergences pertain exclusively to perturbation 
theory and are removed by renormalization of the coupling
constants. Quantities expressed in terms of
physicial (renormalized) couplings show no trace of perturbative
divergences. 

  \subsection{Monotonicity of the relative entropy}

From the expression of the relative entropy as a Legendre transform
(\ref{Legtrans}) we obtain
\begin{equation}
{d\cS\over dz} = {dW\over dz} - {d\over dz}\left(z{dW\over dz}\right) =
- z {d^2W\over dz^2} = - z{d\over dz}\langle I_r\rangle = 
-z (-\langle I_r^2\rangle + \langle I_r\rangle^2).
\end{equation}
Then we have the positive quantity
\begin{equation}
z{d\cS\over dz} = z^2\langle (I_r-\langle I_r\rangle)^2\rangle \geq 0,
\end{equation}
which proves that the relative entropy is a monotonic function of $|z|$.
Moreover, 
\begin{equation}
\langle (I_r-\langle I_r\rangle)^2\rangle = 
l^a\langle (f_a-\langle f_a\rangle)(f_b-\langle f_b\rangle)\rangle\, l^b.
\end{equation}
Hence the matrix 
\begin{equation}
Q_{ab}=\langle (f_a-\langle f_a\rangle)(f_b-\langle f_b\rangle)\rangle =
-{\partial^2 W \over{\partial l^a \partial l^b}}
\end{equation}
is positive definite, implying that $W(l^a)$ is convex. This matrix is
generally defined in probability theory, where it is called the Fisher
information matrix and provides a metric in the space of distributions.

There is a similar convexity property for the relative entropy as
function of its natural variables, the expectation values of composite
fields, $\cS(\langle f_a\rangle)$; since
\begin{equation}
{\partial^2 \cS \over{\partial \langle f_a\rangle \partial \langle
f_b\rangle}} = 
- {\partial l^a \over \partial \langle f_b\rangle} = 
-\left({\partial \langle f_b\rangle \over{\partial l^a}}\right)^{-1} = 
-\left({\partial^2 W \over{\partial l^a \partial l^b}}\right)^{-1} =
Q_{ab}^{-1},
\end{equation}
the matrix of second derivatives of the entropy is the inverse of $Q$
and hence positive definite as well. Summarizing, the key properties of
the entropy are its monotonicity with $z$, therefore, 
any coupling, and its convexity in the space of composite fields
$\langle f_a\rangle$. 

\section{Crossover between field theories}

The situation we consider is when the action $I$ is associated with some
multicritical point and there is crossover to another lower
multicritical point. Crossover signifies a change of the
type of critical behaviour or universality class, with a change of
critical exponents or any other quantity pertaining to it.
Then $I_0$ refers to the critical action at the
first point and $I_r$ contains the relevant coupling constants. The way
in which a crossover occurs is illustrated by the action of
the RG: Its flow will lead away from the first point asymptotically to a
region of the phase diagram of lower dimensionality, where one or more
couplings become irrelevant and dissapear.

We will restrict our considerations in what follows to
scalar $\IZ_2$ symmetric field theories in $2<d<4$ with polynomial
potentials and non symmetry breaking fields.  For illustration, we
will discuss some exact results pertaining to soluble statistical models,
which illuminate the behaviour of the field theories in the same universality
classes.

\subsection{The Gaussian model and the zero to infinite mass crossover}

As an elementary example we take the Gaussian model, given by the action
\begin{equation}
I[\phi]
=\int_{\M}\left\{{1\over2}(\partial\phi)^2+{r\over2}\phi^2\right\}.
\end{equation}
Here the relevant coupling is $r$. Its critical value is $r_c=0$. Hence
the crossover parameter is also $r$ and the end fixed point for
large $r$ is the infinite mass Gaussian model. 

The Gaussian model is soluble in any dimension by direct functional
integration. The lattice version with coupling $\beta=1/T$ yields \cite{Parisi}
\begin{equation}
W[\beta]={1\over2}\int_{-\pi}^{\pi}{d^dk\over{(2\pi)}^d}\,
\ln\left\{1 - 2\beta\,\sum_i^d \cos k_i\right\}
\end{equation}
per site.
The continuum limit is performed by redefining momenta as $k=ap$ and 
considering $W$ per unit volume. Since
$k$ belongs to a Brillouin zone, $-\infty<p<\infty$. It is straight
forward to check that the UV divergences are as announced in the
previous section and cancel in the relative entropy.  

\subsection{Ising universality class}

Let us next consider the Ising model on a rectangular lattice.
For simplicity we will restrict our considerations to equal couplings in the
different directions.  Since the random variables here (the Ising spins) take
discrete values it is natural to consider the absolute entropy. This
model, as is well known, admits an exact solution in two 
dimensions for the partition function \cite{Onsager} which yields a $W$
per site 
\begin{equation}
W[\beta] ={1\over 2}\int_{-\pi}^\pi {d^2k\over{{(2\pi)}^2}}\,
\ln\left[\cosh^2(2\beta)- \sinh(2\beta)(\cos k_x+\cos k_y) \right].
\label{Wps}
\end{equation}
The entropy is then
$$S_a=-\left(W(\beta)-\beta{dW(\beta)\over d\beta}\right)$$
as plotted against $\beta$ in figure 1a.
The monotonicity  of the entropy turns to 
convexity when it is expressed in terms of the internal
energy $U={dW\over d\beta}$ as can be seen in figure 1b.
\vfill\eject
\phantom{a}
\par\vskip-2cm
\def\epsfsize#1#2{.3#1}
\centerline{\epsfbox{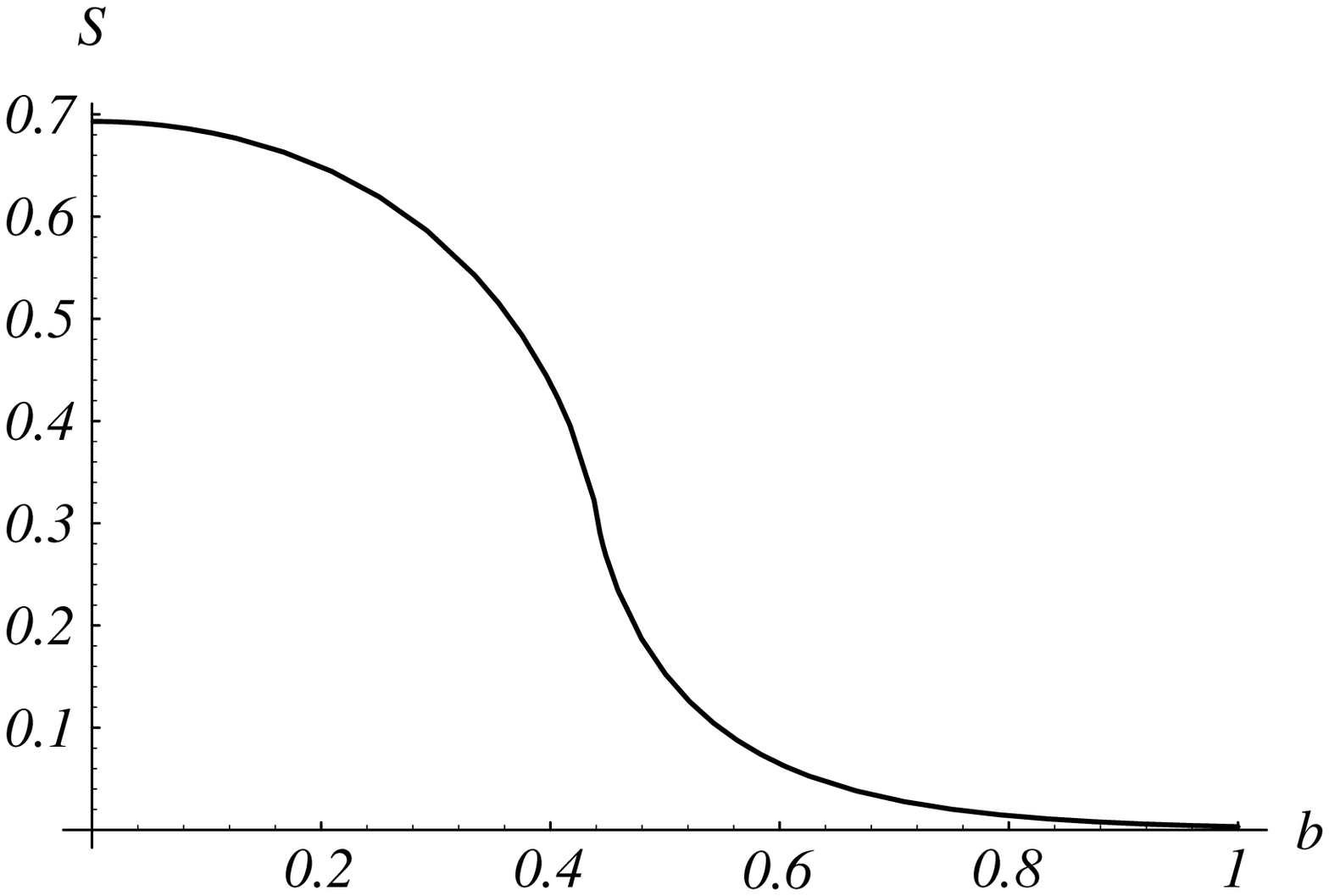}\hspace{1cm}\epsfbox{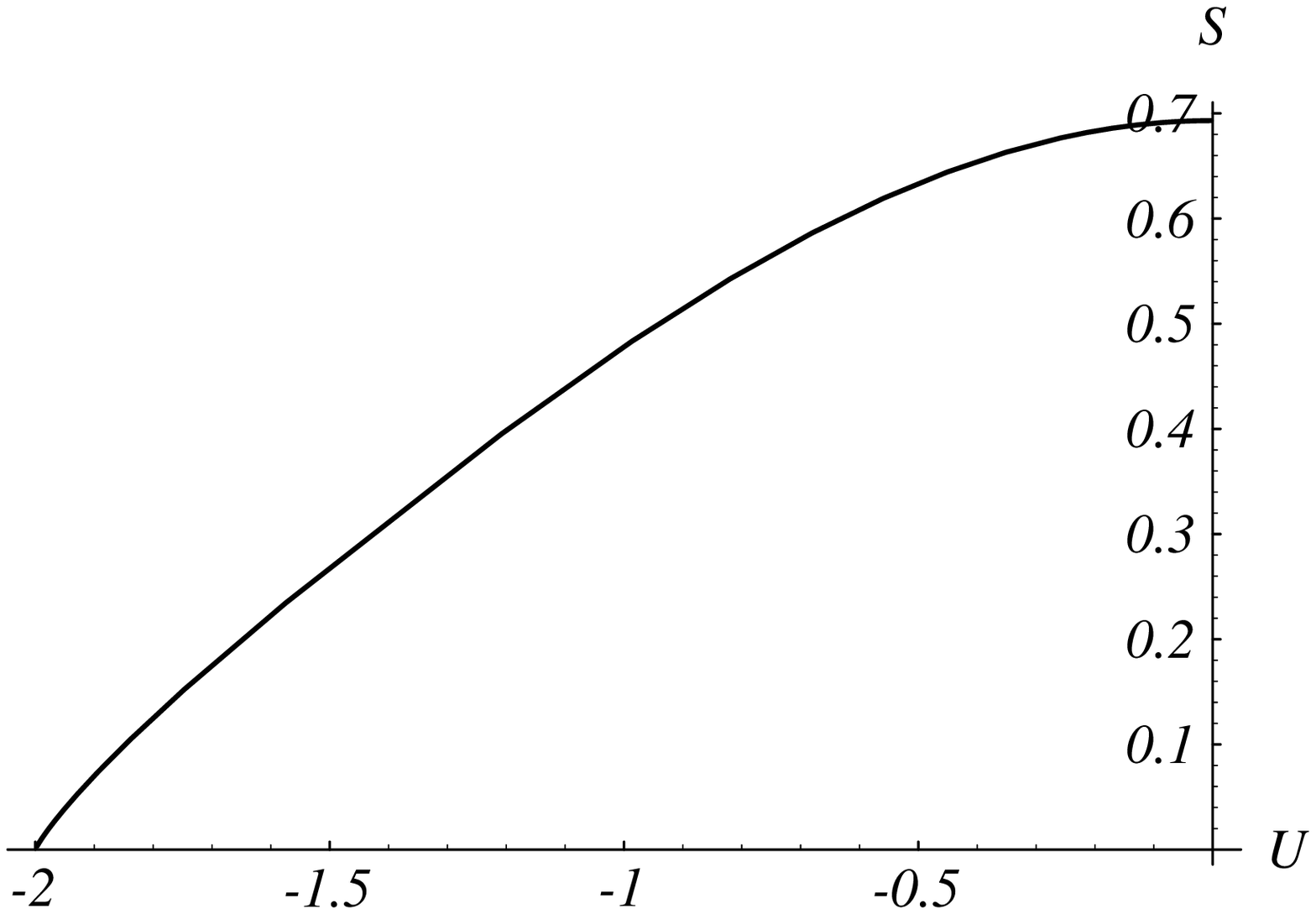}}
\par\vskip-1cm
\begin{center}
{\bf Fig.\ 1a and 1b:} The entropies $S_a(\beta)$ and $S_a(U)$ for the 2d
Ising model
\end{center}
\def\epsfsize#1#2{.3#1}
\centerline{\epsfbox{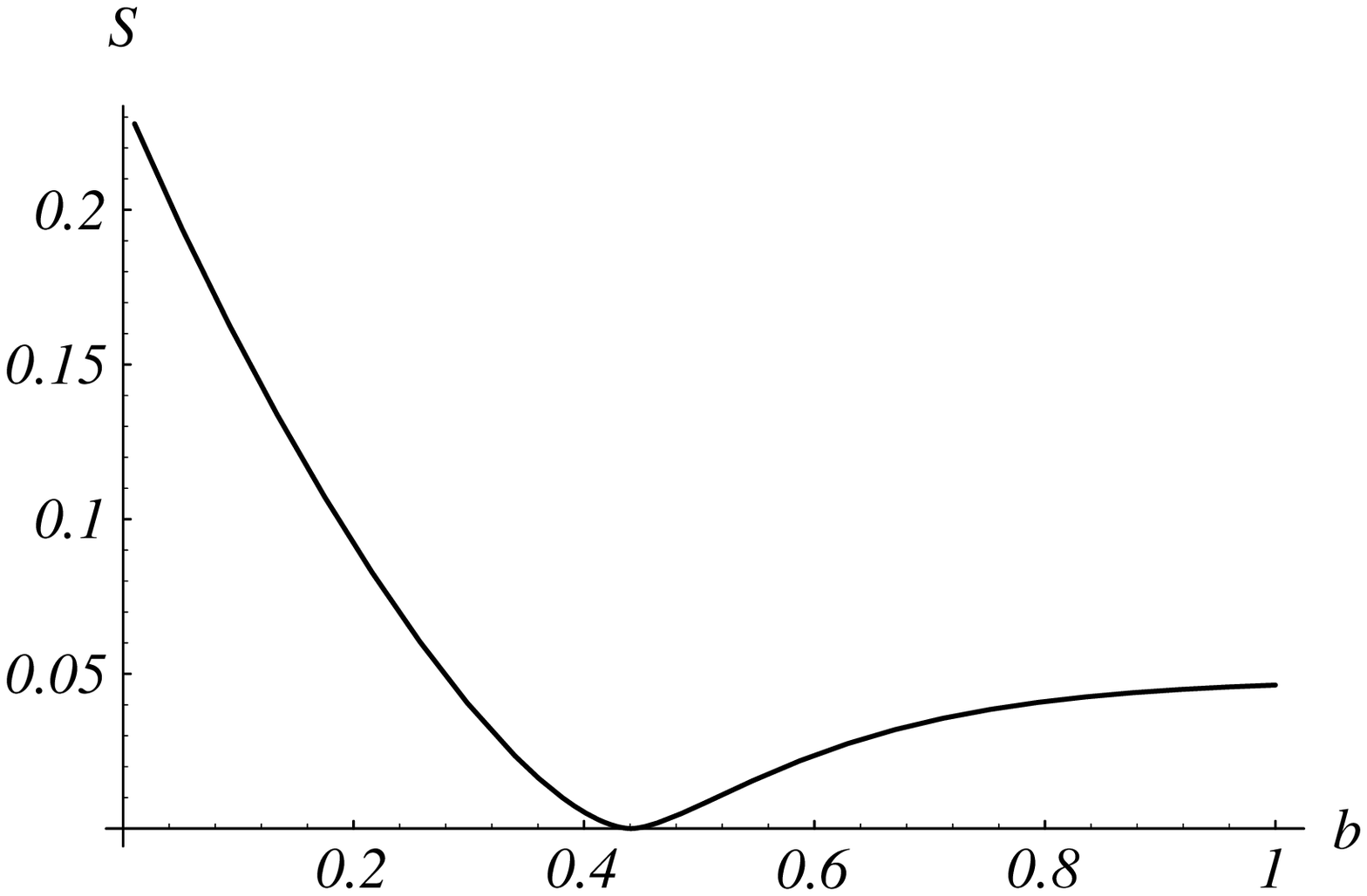}\hspace{1cm}\epsfbox{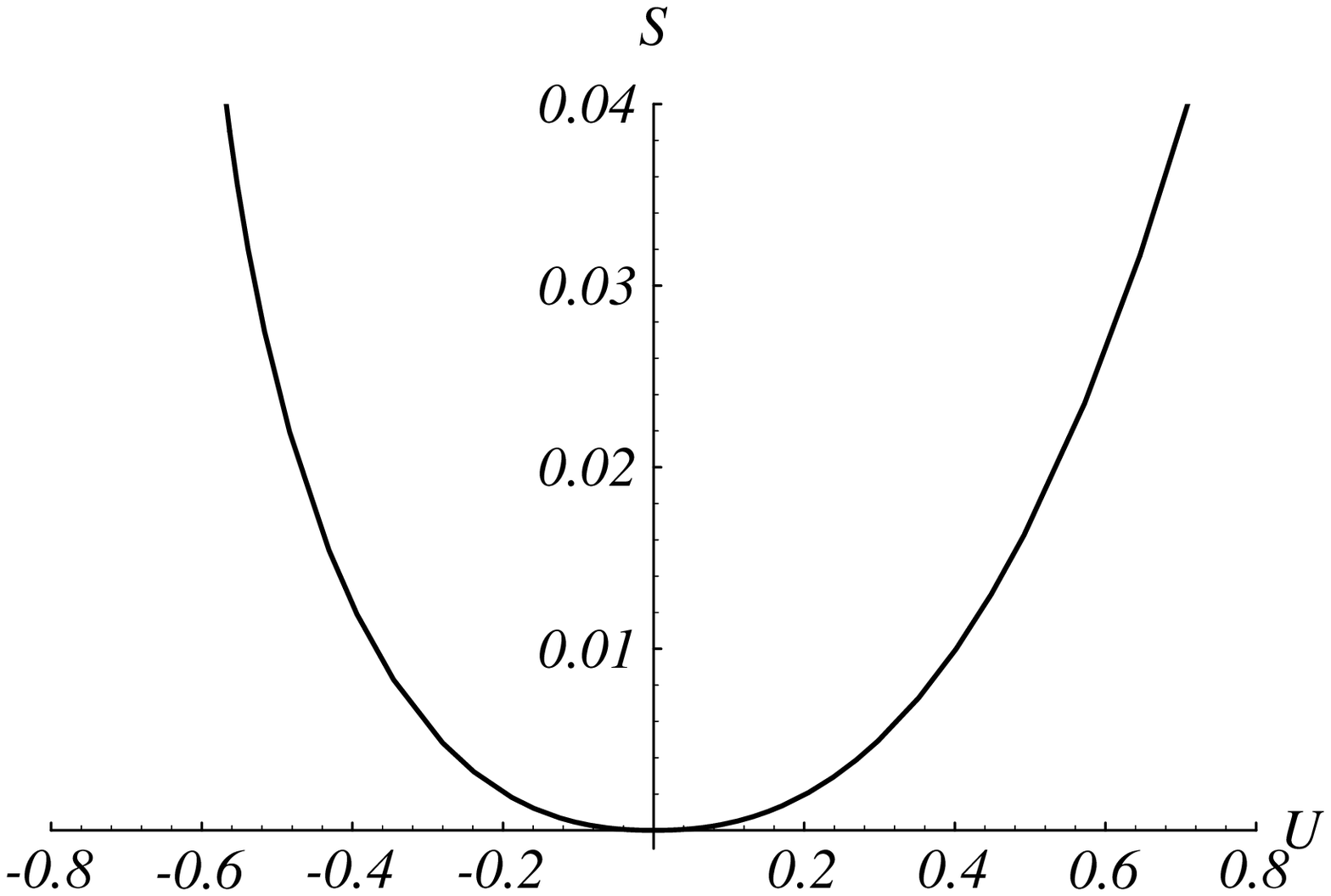}}
\par\vskip-1cm
\begin{center}
{\bf Fig.\ 2a and 2b:} The relative entropies $S(\beta,\beta^*)$ and
$S(U,U^*)$ for the 2d Ising model  
\end{center}
\vskip1cm
Now, of course, we can also consider relative entropy in this setting.
To facilitate
comparison with a field theory it is natural to choose entropy
relative to the critical point (CP) lattice Ising model. This is also natural
since the
critical point is a preferred point in the model. This relative entropy is
given by $$S=W(\beta)-W(\beta^*)-(\beta-\beta^*){dW(\beta)\over d\beta}$$
where $\beta^*={1\over2}\ln(\sqrt{2}+1)\sim 0.4406868$ is the
critical coupling of the Ising model.
We have plotted this in figure 2a. We see that it is a monotonic
increasing function of $|\beta-\beta^*|$ and is zero at the
critical point. In figure 2b we plot this entropy as a function
of the relevant expectation value, the internal energy
$U={dW\over d\beta}$,  and set the origin at  $U^*$,
the internal energy at the critical point. Naturally, the graph
is convex.

In the continuum limit the absolute entropy or $W$ per unit volume are
UV divergent 
but the relative entropy is finite, like in the Gaussian model. 
In fact,  $W$ is similar in both models when
expressed in 
terms of the mass of the particles, although the relation of the mass
with the coupling constant is very different in each case, of course. 

In more than two dimensions the Ising model has not been solved exactly.
Its critical behaviour is in the universality class of a $\phi^4$ field
theory, 
\begin{equation}
I[\phi]
=\int_{\M}\left\{{\alpha\over2}(\partial\phi)^2+{r\over2}\phi^2+{\l\over
4!}\phi^4\right\}.
\end{equation}
We restrict our considerations to $d<4$ where the theory is
super-renormalizable. 
The CP occurs for $r_c$ such that the correlation length is infinite, 
with some arbitrary but fixed value of the bare coupling constant
$\l$. These are the reference values for the relative entropy. 
Since the squared mass $r$ is a function of the Ising coupling
$\beta$, going off from $\beta_c$ induces crossover: This is the
crossover line from 
the Wilson Fisher fixed point to the infinite mass Gaussian fixed point.
Now we can apply the previous monotonicity theorem with $z=r-r_c$.
Since $\l$ is relevant in $d<4$, 
one can place it into the crossover portion of the action. This
provides us with another crossover and in this more complicated 
phase diagram there are in fact two Gaussian fixed points: 
A massless and an infinite mass one,
both associated with $\l=0$ (see the article by Nicoll et al \cite{NCS} for a
description of the total phase diagram).
The crossover between them is that associated with
the Gaussian model as described above.

\subsection{The universality class of the tricritical model}

We consider now models with a tricritical point (TCP), namely, with a phase
diagram where three critical lines meet at a point and three phases
become critical simultaneously. The universality class of those models
is a field theory with a sixth degree coupling, with action
\begin{equation}
I[\phi,\{\l(2)\}]
=\int_{\M}\left\{{1\over2}(\partial\phi)^2
+{r\over2}\phi^2+{\l\over4!}\phi^4+{g\over 6!}\phi^6\right\}
\end{equation}
This field theory is superrenormalizable for $d<3$.
The TCP occurs for $r_{tc}$ and $\l_{tc}$ such that the correlation
lengths of the two order parameters $\phi$ and $\phi^2$ are infinite, 
with some arbitrary but fixed value of the bare coupling constant
$g$. These are the reference values for the relative entropy. 

Now we have various possible crossovers. If $g$ is held fixed the
relevant variables are $t=r-r_{tc}$ and $l=\l-\l_{tc}$. 
First consider the line formed setting $l=0$ and ranging
$t$ from zero to infinity. This is a line leaving the tricritical point and
going to
an infinite mass Gaussian model. 
Similarly we can consider the line $t=0$ and $l$ ranging through different
values. If $\l$ is held at its critical value one sweeps a critical line
and crosses over to the Wilson-Fisher fixed point.
According to the monotonicity theorem we have again that the
relative entropy is a monotonic function in these crossovers.

A more thorough analysis of the structure of the phase diagram pertinent
to these crossovers can be done in terms of the shape of RG trajectories. They
can be easily obtained near the TCP from the linear RG equations as 
$t=c\,l^{\vf}$ for various $c$, with only one parameter given by the
ratio of scaling dimensions of the relevant fields
$\vf = {\Delta_t \over\Delta_l} >1$, called the crossover exponent.
These curves have the property that
they are all tangent to the $t$ axis at the origin and any straight line
$t=a\,l$ intersects them at some finite point,
$l_i={({a/c})}^{1\over \vf-1}$
and $t_i=a l_i$.
For any given $c$ the values of $l_i$ and $t_i$ increase as $a$
decreases and go to infinity as $a \rightarrow 0$. This clearly shows
that the stable fixed point of the flow is on the line at infinity and, in
particular, its projective coordinate is $a=0$.
The point $a=\infty$ on the line at infinity is also fixed but unstable.
In general, as the overall factor $z$ is taken to infinity
we shall hit some point on the
separatrix connecting these two points at infinity.
 
We recognize here a feature of the type of crossover phase diagrams that
we study: In a coordinate system defined by the linear RG (usually
called system of nonlinear scaling fields) near the highest
multicritical point the remaining fixed points are located at infinity. 
To study the crossover, when a FP is at infinity, we need to perform
some kind of compactification of the phase diagram.
Thus, we shall think of the total phase diagram as a compact manifold
containing the maximum number of {\it generic} RG FP.
This point of view is especially sensible regarding the topological
nature of RG flows. Furthermore, thinking
of the RG as just an  ODE indicates
what type of compactification of phase diagrams is adequate:
It is known in the theory of ODEs that the analysis of the
flow at infinity and its possible singularities can be done
by completing the affine space to
projective space.\cite{Lefschetz} This is also appropriate for phase diagrams. 

The projective compactification lends itself to an interesting physical
interpretation. In the tricritical action, for example, we can
consider the set of couplings $r$, $\l$ and $g$ as a system of
homogeneous coordinates in the real projective space ${\IR}P_2$.
Holding $g$ fixed is equivalent to taking the patch of ${\IR}P_2$
appropriate to the TCP, with affine coordinates $r/g$ and $\l/g$.
(We may set $g=1$, say, where we now use dimensionless
couplings, the original $g$, which we now label $g_B$, setting the scale.)
Following the RG trajectories, in the IR limit $g_R$ becomes independent
of its initial value and eventually goes to zero, as illustrated by the
solution of the one-loop RG equations,
$$g_{R}={g_{B}\over
1+a(d)\,g_{B}\,R^{3-d}}$$
with $R$ the IR cutoff and $a(d)$ a dimension dependent factor.
This is the physical reason why the other FP are at infinite distance of
the TCP and one must take a coordinate system with $g=0$ or, in
mathematical terms, choose another coordinate patch in ${\IR}P_2$.

\section{Wilson's RG and $H$ theorem}

A Wilson RG transformation is such
that it eliminates degrees of freedom of short wave length and hence
high energy. Typical examples are decimation or block spin transformations.
It is intuitively clear that their action discards information on
the system and therefore must produce an increase of entropy.
Indeed, as remarked by Ma \cite{Ma} iterating this type
transformation does not constitute a group but rather a semi-group, since
the process cannot be uniquely reversed. In the language of statistical
mechanics we can think of it as an irreversible process.

We illustrate Wilson's RG by a very simple example,
the Gaussian model of subsection 3.1 with cut-off action
\begin{equation}
I={1\over2}\int_{0}^{\Lambda}d^dp\,\,\phi(p)\,\left(p^2+r\right)\,\phi(-p),
\end{equation}
which yields
\begin{equation}
W[z]={1\over2}\int_{0}^{\Lambda}{d^dp\over
{(2\pi)}^d}\,\ln{p^2+r\over\Lambda^2}.
\end{equation}
The corresponding relative entropy 
\begin{equation}
\cS[z]={1\over2}\int_{0}^{\Lambda}{d^dp\over(2\pi)^d}
\left( \ln{p^2+r\over {p^2+r_c}} - {t\over p^2+r}\right)
\end{equation}
is finite when $\Lambda$ goes to infinity and vanishes for $t=0$. The
differential Wilson RG is 
implemented by letting $\Lambda$ run to lower values. Let us see that $S$
is monotonic with $\Lambda$.

We have that 
\begin{equation}
{\partial \cS\over\partial\Lambda} = {\Lambda^{d-1}\over
2^d \pi^{d\over 2}\, \Gamma(d/2)} \left(
\ln{\Lambda^2+r\over {\Lambda^2+r_c}} - {t\over \Lambda^2+r}\right),
\end{equation}
except for an irrelevant constant. With the change of variable $x =
\Lambda^2$, we have to show that the corresponding function of $x$ is
of the same sign everywhere. Then we want $$\ln{x+r\over x+r_c} - {r-r_c\over
x+r}$$ not to change sign. Interestingly, the properties of this
expression are independent of $x$ somehow for if one substitutes in
$\ln\rho - {\rho -1 
\over\rho}$ the value $\rho = {x+r\over x+r_c}$ then one recovers the entire
function. 
Now it is easy to show that $\ln\rho \geq 1 - {1\over\rho}.$ (The
equality holds for $\rho =1$---the critical point.) 

This proof resembles the classical proofs of $H$ theorems.
Boltzmann's $H$ theorem ($H=-S$ of (\ref{entropy})) states that 
the coarse-grained entropy of a rarefied gas, described by its one-particle
statistical distribution 
$f(\bm r, \bm p)$, increases as the gas evolves to its
Maxwell-Boltzmann equilibrium distribution, where it remains constant,
effectively making this evolution an irreversible process.\cite{Balian}
His proof uses the expression for ${df\over dt}$ given by the kinetic
equation that takes into account collisions between particles. However,
the negativity of ${dH\over dt}$ is ultimately due to the positivity of
the function $(x-1)\,\ln x$, in a similar fashion to the
proof in the previous paragraph.

For general far-from-equilibrium processes a description in terms of
differential equations (with respect to time) is not available.
A typical irreversible process is, for example, the one that takes
place when the 
partition which thermally insulates two parts of a container with
different temperature is suddenly removed. The container (the total
system) passes through states which cannot be regarded as
thermodynamical states, that is, described by a small number of
variables, until it reaches a new equilibrium state where the
temperature is well defined and any trace of the initial values in the
two parts is lost. The important point is that, since the 
entropy is a state variable, its increase is independent of the details
of the process and one can assume a quasi-static process for
convenience. 

We can think of the action of Wilson's RG in this manner: typically its
action introduces new couplings and eventually their number is not
bound and the description in terms of the original Hamiltonian is no
longer useful. However, when a FP is approached the number of couplings
reduces again, in fact to a smaller number than the initial one since
some information is necessarily lost: A relevant coupling must become
irrelevant in the process. Here too it may be convenient a description of the
process, a crossover, by quasi-equilibrium states, such as
is provided by the Field Theory picture presented in the previous
section. 
However, to independently establish the increase of entropy we can
ressort to the very general formultion of the $H$ theorem provided by
Jaynes in the form of what has been called the maximum entropy
principle.\cite{Jaynes}

In Jayne's maximum entropy principle $H$ is a function(al)
of the probability distribution of the system that measures the
information available to 
the system and has to be a minimum at equilibrium. To be precise, the
actual probability distribution
is such that it does not contain information other than that implied
by the constraints or boundary conditions imposed at the outset.
The simplest case of the $H$ theorem is when there is no constraint
wherein $H$ is a minimum for a uniform distribution. This is sometimes called
the principle of equiprobability.
This is the case for an isolated system in statistical mechanics:
all the states of a given energy have the same probability
(micro-canonical distribution). Another illustrative example 
is provided by a system thermally coupled
to a heat reservoir at a given temperature where we want to impose
that the average energy takes a particular value.
Minimizing $H$ then yields the canonical distribution.

In general, we may impose constraints on a system with states $X_i$ that
the average values of a set of functions of its state, $f_r(X_i)$,
adopt pre-determined values,
\begin{equation}
\langle f_r\rangle := \sum_i P_i\, f_r(X_i) = \bar f_r,
\end{equation}
with $P_i := P(X_i)$. The maximum entropy formalism leads to
the probability distribution \cite{Tribus}
\begin{equation}
P_i =  Z^{-1}\,\exp {\left(-\sum_r \lambda_r\,f_r(X_i) \right)}.
\end{equation}
The $\lambda_r$ are Lagrange multipliers determined in terms
of $\bar f_r$ through the constraints.
In field theory a state is defined as a field configuration $\phi(x)$.
One can define functionals of the field ${\cal F}_r[\phi(x)]$.
These functionals are usually quasi-local and are called composite
fields. The physical input of a theory can be given in two ways, either by
specifying the microscopic couplings or by specifying the expectation
values of some composite fields, $\langle {\cal F}_r[\phi(x)]\rangle$.
The maximum entropy condition provides an expression
for the probability distribution,
\begin{equation}
P[\phi(x)] = Z^{-1}\,
{\rm exp} {\left(-\sum_r \lambda_r\,{\cal F}_r[\phi(x)] \right)},
\label{PJay}
\end{equation}
and therefore for the action,
\begin{equation}
I = \sum_r \lambda_r\,{\cal F}_r;
\end{equation}
namely, a linear combination of relevant fields with coupling constants
to be determined from the specified $\langle {\cal F}_r\rangle$.
If a constraint is released, namely, the information given by the
expectation value of a composite field is lost, the system
evolves---crosses over---to a more stable state with larger entropy. Its
probability distribution is again of the form (\ref{PJay}) but with one
less term in the sum, the irrelevant field that has dissapeared. 

\section{Conclusions}

We have shown that the relative entropy is the natural
definition of entropy in Field Theory; namely, this relative
entropy is well defined, that is, free 
from divergences, as opposed to the absolute entropy. It is then the
appropriate quantity to study crossover between field theories.

We have established a theorem of monotonicity of the relative entropy
with respect to the coupling constants.
Thus it provides a
natural function which ranks the different critical points in a
model. It grows 
as one descends the hierarchy in the crossovers between scalar field
theories corresponding to different multicritical points.

We have further established that the phase diagrams of the hierarchy of
critical points are associated with a nested sequence of projective
spaces. It is convenient to use coordinates adapted to a particular
phase diagram in the hierarchy. Hence a crossover implies a coordinate
change. The transition from bare to renormalized coordinates provides a
method of compactifying the phase diagram. 

We discussed the action of the Wilson RG and argued that
the relative entropy increases as more degrees of freedom are
integrated out. Although a differential increase may be hard to get in
general due to the proliferation of couplings, Jaynes' formulation of
the $H$ theorem allows us to conclude that the entropy increases globally
in a crossover. 

This study was motivated by the search for a monotonic function
{\em along} the RG trajectories, like the $C$ function given by
Zamolodchikov's theorem in two dimensions. Thus we owe a comment to the
connection with this theorem. The field theoretic relative entropy is a
well defined and useful quantity on its own, as we have shown. Since we
restricted 
the dimension to $d>2$ a direct comparison with the $C$ function is
not possible. Given that the $C$ function is built from correlation data,
unlike the relative entropy, they do not seem related. However, recent
work \cite{Hol} relates the central charge with a particular type of
entropy, the geometrical entropy. Undoubtedly, this connection
needs to be further investigated.

\section*{Acknowledgments}

I am grateful to Denjoe O'Connor for collaboration in the work on
which this report is based and for conversations. I am also grateful to
M. L{\"a}sig, T. Morris and C. Stephens for questions that helped to focus
myself on the importance of the field theoretic relative entropy being
well defined.

 \end{document}